\documentclass[sigconf]{acmart}

\usepackage{amsmath,amsfonts,bm}

\def\eqref#1{equation~\ref{#1}}

\def\1{\bm{1}}

\def\vh{{\bm{h}}}

\def\vm{{\bm{m}}}

\def\mA{{\bm{A}}}

\def\mH{{\bm{H}}}

\def\mP{{\bm{P}}}
\def\mQ{{\bm{Q}}}

\def\mT{{\bm{T}}}

\DeclareMathAlphabet{\mathsfit}{\encodingdefault}{\sfdefault}{m}{sl}
\SetMathAlphabet{\mathsfit}{bold}{\encodingdefault}{\sfdefault}{bx}{n}

\usepackage{hyperref}
\usepackage{url}
\usepackage{graphicx}
\usepackage{amsmath,amsthm}
\usepackage{booktabs}
\usepackage{thmtools,thm-restate}
\usepackage{mathtools}
\usepackage{subfig}
\usepackage{multirow}
\usepackage{colortbl}
\usepackage{enumitem}
\usepackage{color}
\usepackage{xcolor}
\usepackage{makecell}
\usepackage[linesnumbered,boxed,ruled,vlined]{algorithm2e}
\usepackage{listings}
\usepackage{threeparttable}
\usepackage[export]{adjustbox}

\usepackage{flushend}
\usepackage{makecell}
\usepackage[normalem]{ulem} 
\usepackage{bbding}

\AtBeginDocument{%
  }

\copyrightyear{2023}
\acmYear{2023}
\setcopyright{rightsretained}
\acmConference[WWW '23]{Proceedings of the ACM Web Conference 2023}{April 30-May 4, 2023}{Austin, TX, USA}
\acmBooktitle{Proceedings of the ACM Web Conference 2023 (WWW '23), April 30-May 4, 2023, Austin, TX, USA}
\acmDOI{10.1145/3543507.3583472}
\acmISBN{978-1-4503-9416-1/23/04}

\newcommand{\hide}[1]{}

\newcommand{\vpara}[1]{\vspace{0.2cm}\noindent\textbf{#1 }}

\newcommand{\revise}[1]{{#1}}

\usepackage{xspace}

\definecolor{lightgreen}{RGB}{197,224,180}
\definecolor{lightblue}{RGB}{222,235,247}
\definecolor{lightpurple}{RGB}{238,229,241}
\definecolor{lightorg}{RGB}{251,229,214}

\begin{document}


\title{CogDL: A Comprehensive Library for Graph Deep Learning}


\settopmatter{authorsperrow=3}

\author{Yukuo Cen}
\email{cyk20@mails.tsinghua.edu.cn}

\author{Zhenyu Hou}
\email{houzy21@mails.tsinghua.edu.cn}

\author{Yan Wang}
\email{wang-y17@mails.tsinghua.edu.cn}
\affiliation{Tsinghua University\country{China}}

\author{Qibin Chen}
\email{cqb19@mails.tsinghua.edu.cn}

\author{Yizhen Luo}
\email{luoyz18@mails.tsinghua.edu.cn}

\author{Zhongming Yu}
\email{yzm18@mails.tsinghua.edu.cn}
\affiliation{Tsinghua University\country{China}}

\author{Hengrui Zhang}
\email{hengrui-18@mails.tsinghua.edu.cn}

\author{Xingcheng Yao}
\email{yxc18@mails.tsinghua.edu.cn}

\author{Aohan Zeng}
\email{zah22@mails.tsinghua.edu.cn}
\affiliation{Tsinghua University\country{China}}

\author{Shiguang Guo}
\affiliation{Tsinghua University\country{China}}
\email{gsg18@mails.tsinghua.edu.cn}

\author{Yuxiao Dong}
\affiliation{Tsinghua University\country{China}}
\email{yuxiaod@tsinghua.edu.cn}

\author{Yang Yang}
\affiliation{Zhejiang University\country{China}}
\email{yangya@zju.edu.cn}

\author{Peng Zhang}
\affiliation{Zhipu AI\country{China}}
\email{zpjumper@gmail.com}

\author{Guohao Dai}
\affiliation{Shanghai Jiao Tong University\country{China}}
\email{daiguohao@sjtu.edu.cn}

\author{Yu Wang}
\affiliation{Tsinghua University\country{China}}
\email{yu-wang@tsinghua.edu.cn}

\author{Chang Zhou}
\affiliation{Alibaba Group\country{China}}
\email{ericzhou.zc@alibaba-inc.com}

\author{Hongxia Yang}
\affiliation{Alibaba Group\country{China}}
\email{yang.yhx@alibaba-inc.com}

\author{Jie Tang*}
\affiliation{Tsinghua University\country{China}}
\email{jietang@tsinghua.edu.cn}

\renewcommand{\authors}{Yukuo Cen, Zhenyu Hou, Yan Wang, Qibin Chen, Yizhen Luo, Zhongming Yu, Hengrui Zhang, Xingcheng Yao, Aohan Zeng, Shiguang Guo, Yuxiao Dong, Yang Yang, Peng Zhang, Guohao Dai, Yu Wang, Chang Zhou, Hongxia Yang, and Jie Tang}

\renewcommand{\shortauthors}{Cen et al.}

\begin{abstract}

\renewcommand{\thefootnote}{\fnsymbol{footnote}}
\footnotetext[1]{Jie Tang is the corresponding author.}
\renewcommand{\thefootnote}{\arabic{footnote}}

Graph neural networks (GNNs) have attracted tremendous attention from the graph learning community in recent years. 
It has been widely adopted in various real-world applications from diverse domains, such as social networks and biological graphs. 
The research and applications of graph deep learning present new challenges, including the sparse nature of graph data, complicated training of GNNs, and non-standard evaluation of graph tasks. 
To tackle the issues, we present CogDL\footnote{The code is available at: \url{https://github.com/THUDM/cogdl}}, a \textbf{\textit{co}}mprehensive library for \textbf{\textit{g}}raph \textbf{\textit{d}}eep \textbf{\textit{l}}earning that allows researchers and practitioners to conduct experiments, compare methods, and build applications with ease and efficiency. 
In CogDL, we propose a unified design for the training and evaluation of GNN models for various graph tasks, making it unique among existing graph learning libraries. 
By utilizing this unified trainer, CogDL can optimize the GNN training loop with several training techniques, such as mixed precision training. 
Moreover, we develop efficient sparse operators for CogDL, enabling it to become the most competitive graph library for efficiency. 
Another important CogDL feature is its focus on ease of use with the aim of facilitating open and reproducible research of graph learning. 
We leverage CogDL to report and maintain benchmark results on fundamental graph tasks, which can be reproduced and directly used by the community. 

\end{abstract}

\ccsdesc[500]{Software and its engineering~Software libraries and repositories}
\ccsdesc[500]{Mathematics of computing~Graph algorithms}
\ccsdesc[300]{Computing methodologies~Learning latent representations}
\keywords{graph deep learning, graph neural networks, graph library.}

\maketitle

\section{Introduction}
\label{sec:intro}

Graph-structured data have been widely utilized in many real-world scenarios. 
Inspired by recent trends of representation learning on computer vision (CV) and natural language processing (NLP), graph neural networks (GNNs)~\cite{gori2005new,scarselli2008graph} are proposed to apply neural architectures to perform message passing over graph structures. 
For example, 
graph convolutional networks (GCNs)~\cite{kipf2016semi} simplify the spectral graph convolutions with first-order approximation and utilize a layer-wise propagation rule of graph convolution. 
GraphSAGE~\cite{hamilton2017inductive} adopts a sampling-based method to generate node embeddings for large graphs, which can be applied to the inductive setting. 
Recent effort in GNNs has been focused on theoretical understanding~\cite{xu2019gin}, generalization~\cite{feng2020grand}, self-supervised learning~\cite{velickovic2019dgi,Hu:KDD20GPT}, and capacity to handle web-scale applications~\cite{ying2018graph}. 
To date, GNNs have achieved impressive performance on various graph machine learning tasks in diverse domains~\cite{hu2020open}.

\lstdefinestyle{interfaces}{
  float=tp,
  floatplacement=tbp,
  abovecaptionskip=-5pt
}

With the prosperity of GNNs, the evaluation and fair comparison of GNN models can still be a concerning issue. 
Although there are enormous research works on graph machine learning, we observe that different papers may use their evaluation settings for the same graph dataset, making results indistinguishable. For example, as a widely used dataset, Cora~\cite{sen2008collective}, some papers use the ``standard'' semi-supervised splits following Planetoid~\cite{yang2016revisiting}, while others adopt random splits or fully-supervised splits. Reported results of a GNN model on the same dataset may differ in various papers, making it challenging to compare the performance of different GNN methods~\cite{shchur2018pitfalls, errica2019fair, dwivedi2020benchmarking, hu2020open}. 

Moreover, the graph-structured data is fundamentally different from natural language or images, which can be easily formatted as \texttt{Tensor}  and thus straightforwardly processed by GPUs. 
The non-Euclidean and sparse nature of graph data requires 
a better storage format for efficient computation, especially on GPUs. 
However, current deep learning frameworks do not well support the sparse computation of graph-structure data. 
For example, the PyTorch API of sparse tensors is limited and inefficient for graph operations. 
To bridge the gap, several open-source efforts have been made to develop dedicated libraries for the efficient development of graph representation learning research. 
For example, PyTorch Geometric (PyG)~\cite{fey2019fast} is a graph learning library built upon PyTorch~\cite{paszke2019pytorch} to easily write and train GNNs for various applications. 
Deep Graph Library (DGL)~\cite{wang2019deep} is an efficient and scalable package for deep learning on graphs, which provides several APIs allowing arbitrary message-passing computation over large-scale graphs. 
However, these popular graph libraries mainly focus on implementing basic GNN operators and 
do not take the whole training and evaluation of GNNs into consideration.  
The end-to-end performance of training GNNs could still be improved.

\vpara{Present work: CogDL.}
In this paper, we introduce CogDL, a \textbf{\textit{co}}mprehensive library for \textbf{\textit{g}}raph \textbf{\textit{d}}eep \textbf{\textit{l}}earning that allows researchers, practitioners, and developers to train and evaluate GNN models under different graph tasks with ease and efficiency. 
The advantage of CogDL lies in its design of a unified GNN trainer, standard evaluation, and modular wrappers.  
Specifically, CogDL unifies 
the training and evaluation of graph neural network (GNN) models, making it different from other graph learning libraries. 
In addition, it is also equipped with decoupled modules that can be flexibly plugged for training GNNs, including \texttt{ModelWrapper} and \texttt{DataWrapper}. 
CogDL provides built-in GNN-specific wrappers, which could not be offered by other libraries. 

The goal of CogDL is to accelerate research and applications of graph deep learning. 
Based on the design of the unified trainer and decoupled modules, CogDL users can instantiate a GNN trainer  by \texttt{trainer = Trainer(...)} and then call \texttt{trainer.run(...)} to conduct experiments easily, which is not supported by existing graph libraries. 
We also provide a high-level API (i.e., \texttt{experiment}) on top of \texttt{Trainer} for one-line usage. 
Using the unified trainer not only saves users' time in writing additional code but also gives users the opportunity to enable many built-in training techniques, such as mixed precision training. 
For example, users only need to set \texttt{Trainer(fp16=True)} to enable the feature of mixed precision training without any modification. 
Furthermore, CogDL provides as many tasks with standard evaluation as possible for users (as shown in Table~\ref{tab:all_tasks}), 
and maintains experimental results on these fundamental graph tasks with recorded hyperparameter configurations. 
We summarize the characteristics of CogDL as follows:

\noindent
\begin{lstlisting}[style=interfaces,language=Python,breaklines=true,frame=single,numbers=none,caption=One-line code to conduct experiments.,label={lst:one_line}]
from cogdl import experiment
experiment(dataset=["cora","citeseer"], model=["gcn","gat","gcnii"])
\end{lstlisting}

\begin{itemize}[leftmargin=*]
    \item \textbf{Ease of use}. We provide simple APIs in CogDL such that users only need to write one line of code to train and evaluate any graph representation learning methods. CogDL develops a unified trainer with decoupled modules for users to train GNNs easily. Based on this unique design, CogDL can provide extensive features such as hyperparameter optimization, efficient training techniques, and experiment management. 
    \item \textbf{End-to-end performance}. Efficiency is one of the most significant characteristics of CogDL. CogDL develops well-optimized sparse kernel operators to speed up the training of GNN models. 
    For example, these efficient sparse operators enable CogDL to achieve about $2\times$ speedups on the 2-layer GCN~\cite{kipf2016semi} and GAT~\cite{Velickovic:18GAT} models compared with PyG and DGL. 
    \item \textbf{Reproducibility}. Users can quickly conduct the experiments to reproduce the results based on CogDL, as illustrated in Listing~\ref{lst:one_line}. CogDL implements 70+ models and collects 70+ datasets for 10 fundamental graph tasks, as shown in Table~\ref{tab:all_tasks}, facilitating open, robust, and reproducible deep learning research on graphs. 
\end{itemize}

\vpara{Relevance to Web.}
GNNs have been widely used in web research and applications, such as social influence prediction~\cite{qiu2018deepinf,piao2021predicting}, social recommendation~\cite{fan2019graph}, and web advertising~\cite{yang2022wtagraph}. 
However, there still remain several challenges to applying powerful GNN models to real web applications. Practitioners need to not only choose an appropriate GNN but also train the GNNs on large-scale graphs. To mitigate the issues, CogDL provides unique advantages for web research and applications. 
First, the well-optimized sparse operators can speed up the process of new research and applications. 
Besides, the unified trainer with built-in $\texttt{wrappers}$ can be directly utilized to conduct experiments with ease and efficiency. 
In addition, CogDL provides the benchmarking results of GNN methods for all kinds of graph tasks, which could help web-related users to choose appropriate methods. 
Finally, we have successfully applied CogDL to a real-world academic system---AMiner. More details about the application can be found in Appendix~\ref{sec:app}.

\section{The CogDL Framework}
\label{sec:framework}

\begin{table*}
\centering
\caption{All supported tasks, datasets, and models in CogDL. To date, CogDL integrates 10 important graph tasks with the standard evaluation of 70+ datasets. Besides, CogDL implements 70+ graph representation methods, which can be directly used by users. Note that some datasets and methods can be used in several tasks. }
\begin{tabular}{lll}
\toprule
Tasks                             & Datasets                                  & Models                                                                                                                        \\
\midrule
Network Embedding                 & \begin{tabular}[c]{@{}l@{}}PPI, Wikipedia, Blogcatalog, \\DBLP, Flickr, YouTube\end{tabular} & \begin{tabular}[c]{@{}l@{}}DeepWalk~\cite{perozzi2014deepwalk}, LINE~\cite{tang2015line}, node2vec~\cite{grover2016node2vec}, \\ProNE~\cite{zhang2019prone}, NetMF~\cite{qiu2018network}, NetSMF~\cite{qiu2019netsmf}, \\SDNE~\cite{wang2016structural}, Spectral~\cite{tang2011leveraging}, DNGR~\cite{cao2016dngr}, \\GraRep~\cite{cao2015grarep}, HOPE~\cite{ou2016asymmetric}, AutoProNE~\cite{hou2021automated} \end{tabular}  \\
\midrule
Node Classification               & \begin{tabular}[c]{@{}l@{}}Cora, Citeseer, Pubmed, \\ Chameleon, Cornell, Film, \\ Squirrel, Texas, Wisconsin, \\ PPI, Flickr, Reddit, Amazon, Yelp, \\ GitHub, Elliptic, Yelpchi, Alpha, \\ Clothing, Electronics, Wiki, Weibo, \\ BGP, SSN, ogbn-arxiv, ogbn-products, \\ogbn-proteins, ogbn-papers100M, \\ \end{tabular}                    &          \begin{tabular}[c]{@{}l@{}} Chebyshev~\cite{defferrard2016chebynet}, GCN~\cite{kipf2016semi}, GAT~\cite{Velickovic:18GAT}, \\GraphSAGE~\cite{hamilton2017inductive}, APPNP~\cite{klicpera2019appnp}, DGI~\cite{velickovic2019dgi}, \\GCNII~\cite{chen2020gcnii}, MVGRL~\cite{hassani2020mvgrl}, GRAND~\cite{feng2020grand}, \\DropEdge~\cite{rong2019dropedge}, Graph-Unet~\cite{gao2019graph}, GDC~\cite{klicpera2019diffusion}, \\PPRGo~\cite{bojchevski2020scaling}, GraphSAINT~\cite{graphsaint-iclr20}, GraphMix~\cite{verma2021graphmix},\\ SIGN~\cite{sign_icml_grl2020}, MixHop~\cite{abu2019mixhop}, GRACE~\cite{zhu2020deep}, \\DeeperGCN~\cite{li2020deepergcn}, DisenGCN~\cite{ma2019disentangled}, SGC~\cite{sgc}, \\M3S~\cite{sun2020multi}, C\&S~\cite{huang2020combining}, SAGN~\cite{sun2021scalable}, RevGNN~\cite{li2021training}, \\ Cluster-GCN~\cite{chiang2019cluster}, InfoMotif~\cite{sankar2020beyond}, GraphMAE~\cite{hou2022graphmae} \end{tabular}                                                                                                                 \\
\midrule
Graph Classification              & \begin{tabular}[c]{@{}l@{}}MUTAG, PTC, NCI1, PROTEINS, \\IMDB-B, IMDB-M, COLLAB, REDDIT-B, \\ REDDIT-M5k, REDDIT-M12k, NCI09, \\ ENZYMES, ogbg-molhiv, ogbg-molpcba, \\ ogbg-ppa, ogbg-code \\ \end{tabular}                          & \begin{tabular}[c]{@{}l@{}}DGK~\cite{yanardag2015dgk}, graph2vec~\cite{narayanan2017graph2vec}, GIN~\cite{xu2019gin}, \\PATCHY\_SAN~\cite{niepert2016learning}, DiffPool~\cite{ying2018diffpool}, \\InfoGraph~\cite{sun2019infograph}, DGCNN~\cite{wang2019dgcnn}, SortPool~\cite{zhang2018sortpool} \end{tabular}              \\
\midrule
Heterogeneous Node Classification & DBLP, ACM, IMDB                           & \begin{tabular}[c]{@{}l@{}}Simple-HGN~\cite{lv2021we}, GTN~\cite{yun2019gtn}, HAN~\cite{wang2019han}, \\PTE~\cite{tang2015pte}, Metapath2vec~\cite{dong2017metapath2vec}, Hin2vec~\cite{fu2017hin2vec}\end{tabular}                                                                                                  \\
\midrule
Link Prediction                   & \begin{tabular}[c]{@{}l@{}}Cora, Citeseer, Pubmed, ogbl-ppa, \\ogbl-ddi, ogbl-collab, ogbl-citation2 \end{tabular}                    & \begin{tabular}[c]{@{}l@{}}GAE~\cite{kipf2016variational}, VGAE~\cite{kipf2016variational}, \\ network embedding methods\end{tabular}                                                                                                      \\
\midrule
Knowledge Graph Completion        & FB13, FB15k, FB15k-237, WN18, WN18RR                   & \begin{tabular}[c]{@{}l@{}}TransE~\cite{bordes2013translating}, DistMult~\cite{Yang2015EmbeddingEA}, ComplEx~\cite{trouillon2016complex},\\ RotatE~\cite{sun2019rotate}, RGCN~\cite{schlichtkrull2018modeling}, CompGCN~\cite{vashishth2019compositionbased}\end{tabular}                                                                             \\
\midrule
Graph Clustering                  & Cora, Citeseer, Pubmed                                          & GAE~\cite{kipf2016variational}, VGAE~\cite{kipf2016variational}, AGC~\cite{zhang2019attributed}, DAEGC~\cite{wang2019attributed}                            \\
\midrule
Graph Attacks and Defenses & \begin{tabular}[c]{@{}l@{}}grb-cora, grb-citeseer, grb-flickr, \\grb-reddit, grb-aminer\end{tabular} & \begin{tabular}[c]{@{}l@{}}GNN-SVD~\cite{entezari2020all}, GNNGuard~\cite{zhang2020gnnguard}, RobustGCN~\cite{zhu2019robust}, \\RND~\cite{zugner2018adversarial}, TDGIA~\cite{zou2021tdgia}, FGSM~\cite{goodfellow2015explaining}, PGD~\cite{madry2018towards}, \\SPEIT~\cite{zheng2020kdd}, DICE~\cite{waniek2018hiding}, FGA~\cite{chen2018fast}, FLIP~\cite{bojchevski2019adversarial}, \\NEA~\cite{bojchevski2019adversarial}, STACK~\cite{xu2022blindfolded} \end{tabular} \\
\midrule
Multiplex Link Prediction         & Amazon, YouTube, Twitter                  & GATNE~\cite{cen2019representation}, network embedding methods                                                                                                                    \\
\midrule
Top-k Similarity Search & KDD-ICDM, SIGIR-CIKM, SIGMOD-ICDE & GCC~\cite{qiu2020gcc}\\
\bottomrule
\end{tabular}
\label{tab:all_tasks}
\end{table*}

In this section, we present the unified design of the CogDL library, 
which 
is built on PyTorch~\cite{paszke2019pytorch}, a well-known deep learning library. 
Based on PyTorch, CogDL provides efficient implementations of many models and our designed sparse operators. To date, it supports 10 important graph tasks, such as node classification. 
Table~\ref{tab:all_tasks} lists all supported tasks, datasets, and models in the CogDL. 
All the models and datasets are compatible with the unified trainer for conducting experiments on different graph tasks.

\subsection{Unified Trainer}
CogDL provides a unified trainer for GNN models, which takes over the entire loop of the training process. The unified trainer, which contains much engineering code, is implemented flexibly to cover arbitrary GNN training settings. 
We design four decoupled modules for the GNN training, including \texttt{Model}, \texttt{ModelWrapper}, \texttt{Dataset}, \texttt{DataWrapper}. The \texttt{ModelWrapper} is for the training and testing steps, while the \texttt{DataWrapper} is designed to construct data loaders used by \texttt{ModelWrapper}. 
The main contributions of most GNN papers mainly lie in three modules except \texttt{Dataset}. 
For example, the GCN paper trains the GCN model under the (semi-)supervised and full-graph setting, while the DGI paper trains the GCN model by maximizing local-global mutual information. 
The training method of the DGI is considered as a model wrapper named \texttt{dgi\_mw}, which could be used for other scenarios. 
Our unified design also supports the precomputation-based GNN models, such as SIGN~\cite{sign_icml_grl2020} and SAGN~\cite{sun2021scalable}. 

Based on the design of the unified trainer and decoupled modules, we could do arbitrary combinations of models, model wrappers, and data wrappers. For example, if we want to apply the DGI, a representative graph self-supervised learning method, to large-scale datasets, all we need is to substitute the full-graph data wrapper with the neighbor-sampling or clustering data wrappers without additional modifications. 
If we design a new GNN architecture for a specific task, we only need to write essential PyTorch-style code for the model. The rest could be automatically handled by CogDL by specifying the model wrapper and the data wrapper. 
We could quickly conduct experiments for the model using the trainer via \texttt{trainer = Trainer(epochs,...)} and \texttt{trainer.run(...)}. 
Moreover, based on the unified trainer, CogDL provides native support for many useful features, including hyperparameter optimization, efficient training techniques, and experiment management without any modification to the model implementation. 

We introduce efficient techniques of GNN training, which credit to the unified trainer of CogDL. These features could be enabled through \texttt{fp16=True} or \texttt{actnn=True} in \texttt{Trainer} API.

\vpara{Mixed Precision Training.}
CogDL supports mixed precision training, which is a popular technique to relieve the GPU memory and speed up the training process. PyTorch provides a convenient method for mixed precision in \texttt{torch.cuda.amp}, which integrates NVIDIA apex. 
For example, we conduct experiments to test the performance of mixed precision training. 

\vpara{Activation Compression Training.}
With the emergence of deep GNNs, the activation footprints occupy more and more GPU memory. 
And we need to store the activation output of each layer to compute the gradient in the backward pass, which costs much GPU memory in the training step. 
We extend the activation compression training technique~\cite{chen2021actnn,liu2022gact} to GNNs. 
The main advantage of activation compression training is that the GPU memory of training could dramatically decrease.

Based on the provided trainer, CogDL also supports the following features for conducting experiments with ease. 

\vpara{Hyperparameter Optimization.}
Hyperparameter optimization (HPO) is an important feature for GNN libraries since current GNN models utilize more hyperparameters than before. 
We integrate a popular library, Optuna~\cite{akiba2019optuna}, into CogDL to enable HPO. Optuna is a fast AutoML library based on Bayesian optimization. CogDL implements hyperparameter search and even model search for graph learning. 
The key of the HPO is to define a search space. The search space will be automatically utilized by CogDL to start searching and output the best results. 
The usage of HPO can be found in Appendix~\ref{subsec:basic_usage}.

\vpara{Experiment Management.}
Experiment management is crucial for training deep learning models, which could be utilized by researchers and developers for debugging and designing new blocks. CogDL integrates TensorBoard and WandB~\cite{wandb} for experiment management. 
TensorBoard is a common visualization toolkit for tracking metrics like loss and accuracy. 
WandB is a central dashboard to keep track of the experimental status. 
The experiment management could be enabled easily by assigning the \texttt{logger} argument.

\subsection{Efficient Sparse Operators}
\label{sec:efficiency}

We introduce the well-optimized sparse operators of CogDL, including SpMM-like operators, multi-head SpMM, and edge-wise softmax, which are developed and implemented in CogDL for GNN models. 
We first introduce the abstraction of sparse operators in CogDL and its corresponding optimization. 
The General Sparse Matrix-Matrix multiplication (\textit{GSpMM}) operator is widely used in most GNNs.
The reason is that many GNNs apply a general aggregation operation for each node from its neighbors: 

\begin{equation} \label{formula:1}
\centering
\vh_u^{(l+1)}=\phi\left(\left\{\psi\left(\vh_v^{(l)},\vm_e\right)\mid (u,e, v)\in E\right\} \right)
\end{equation}
 
\noindent where $(u,e,v)$ denotes an edge $e$ from node $u$ to node $v$ with a message $\vm_e$, and $\vh_u^{(l)}$ is the hidden representation of node $u$ of layer $l$. When the reduce operation $\phi$ is a summation, and the compute operation $\psi$ is a multiplication, such an operation can be described as an SpMM operator $\mH^{(l+1)} \leftarrow \mA \mH^{(l)}$,
where the sparse matrix $\mA$ represents the adjacency matrix of the input graph $G$, and each row of $\mH^{(l)}$ represents a representation vector of nodes (e.g., $\vh_u^{(l)}$ and $\vh_v^{(l)}$). 

We observe that in existing graph neural network libraries such as PyG \cite{fey2019fast}, or DGL \cite{wang2019deep}, their backend support is not tailored to GNN. In PyG, the message-passing flow is built by torch-scatter, which is borrowed from the design
principles of traditional graph processing. In DGL, its GSpMM function leverage APIs from cuSPARSE \cite{naumov2010cusparse}, which is a vendor library that mainly targets for scientific computing of sparse matrices. 
Another observation is that many previous GPU-based SpMM works fail to be aware of the adaptability of workload balance in graphs. For example, GE-SpMM~\cite{huang2020ge-spmm} and Sputnik~\cite{gale2020sparse} are row-split SpMM methods that could not efficiently support graphs that follow a power-law distribution. 
\revise{CogDL is fully aware of the work-load balance problem compared to PyG/DGL.}
We observe that the algorithm of neighbor group~\cite{bridgethegapgnn, gnnadvisor} could better slice the large nodes in the graph and thus has better efficiency when facing power-law distribution graphs. However, some graphs may naturally conform to node load balancing, such as the k-sampled graph \cite{hamilton2017inductive}, so it is better to use the row-split backend directly. 

To tackle the following problems, 
we propose a new balance abstraction that not only exploits the data of the graph itself but also effectively accelerates our backend. As shown in Listing~\ref{lst:wlb}, our workload balance handler would be generated by a data-aware balancer, which could indicate whether to use a row-split or neighbor group algorithm. Empirically, when the degrees of the majority (90\%) of the nodes are close to the average degree, then we would choose row-split as our backend since it is suitable for graphs that are naturally node balanced. Note that our workload balancer is a preprocessing method since it would generate the auxiliary arrays that are needed by the GSpMM backend. The Listing~\ref{lst:wlb} demonstrates that our GSpMM's inputs include a work balance handler, graph data, and input features. 
Users could choose the reduce and compute operator, as illustrated in Equation~\ref{formula:1}. 

\noindent
\begin{lstlisting}[language=Python,breaklines=true,frame=single,numbers=none,caption=workload balance abstraction in CogDL backend,label={lst:wlb}]
workload_balance = balancer(graph) # preprocessing
out_feat = GSpMM(workload_balance, graph, in_feat, reduce_op, compute_op)
\end{lstlisting}

In our GSpMM design, we implement the warp-based thread mapping algorithm, which would bind different neighbor groups or split rows to different warps. This allows us to assign multiple adjacent rows or neighbor groups into a single CUDA block. Since the same processor shares an L1 cache, adjacent rows or neighbor groups can reduce the L1 loaded data to improve locality. 
\revise{Multi-head SpMM is used to compute the aggregation procedure of GNN models with multi-head attention mechanisms (such as GAT), which means each edge has a different edge weight for each attention head; therefore, multiple SpMM operators are needed. PyG and DGL use multiple SpMM with broadcasting to implement it.}
In our multi-head SpMM kernel, we apply the memory coalescing technique by allocating consecutive warps along the dimension of the head. As different heads share the same location of non-zero elements, this data will be cached by the streaming multiprocessors and reused among different heads, which also saves bandwidth. 

When considering a GAT model, we also need a sampled dense-dense matrix multiplication (SDDMM) operator $\mT=\mA \odot (\mP \mQ^T)$, where $\mT$ and $\mA$ are the sparse matrices, $\mP$ and $\mQ$ are two dense matrices, $\odot$ is element-wise matrix multiplication. The SDDMM operator is used for back-propagating the gradients to the sparse adjacency matrix since the adjacency matrix of the GAT model is computed by the attention mechanism. 
CogDL further speeds up multi-head graph attention by optimizing Edge-wise Softmax, which is defined by 
$\bm{\alpha}'_{uv} = \exp(\bm{\alpha}_{uv}) / (\sum_{v\in \mathcal{N}(u)} \exp(\bm{\alpha}_{uv}))$,
where $\bm{\alpha}_{uv}$ is the attention score between nodes $u$ and $v$. 

In our design of the Edge-wise Softmax kernel, we use the method of warp level intrinsic provided by NVIDIA, a typical method that could accelerate the scan and reduce operation. To prevent the spillover of the exponent, we first apply the scan to find the max value of edge weight and then subtract this maximum from each value. After that, we compute each value by applying the exponent function and reduce all the values within the warp to acquire the sum of all the exponent values. As far as we know, CogDL is the first framework that provides a GPU-based fused Edge-wise Softmax kernel. Therefore, we do not provide the results of the experiment with PyG and DGL.

\begin{table}[t!]
\centering
\caption{Performance of mixed precision training.}
\begin{tabular}{lcccc}
\toprule
         & \multirow{2}{*}{Memory} & \multirow{2}{*}{Accuracy} & \multicolumn{2}{c}{Training Speed}  \\
\cmidrule{4-5}         &          &               & 2080 Ti   & 3090                    \\
\midrule
w/o fp16 & 5,567 MB   & 95.44             & 2.20 it/s & 3.93 it/s               \\
with fp16  & 4,046 MB  & 95.35               & 3.17 it/s & 7.97 it/s               \\
\bottomrule
\end{tabular}
\label{tab:mixed_precision}
\end{table}

\begin{table}[t!]
\centering
\caption{Accuracy (\%) and activation memory (MB) of 2-layer GCNs on three datasets.}
\begin{tabular}{lllll}
\toprule
       & origin (32bit) & act (4bit) & act (3bit) & act (2bit)  \\
\midrule
Flickr & 51.17 (288)    & 51.08 (37)   & 51.14 (26)   & 51.20 (18)    \\
Reddit & 95.33 (1532)   & 95.32 (194)  & 95.31 (158)  & 95.34 (112)   \\
Yelp   & 39.86 (4963)   & 40.06 (773)  & 40.21 (665)  & 39.89 (551)   \\
\bottomrule
\end{tabular}
\label{tab:actnn}
\end{table}

\section{Evaluation}
\label{sec:evaluation}

In this section, we conduct extensive experiments to evaluate the framework design and characteristics of CogDL. 

\subsection{Training Techniques}
For mixed precision training, we run a 2-layer GCN model with 2048 hidden units on the Reddit dataset. From Table~\ref{tab:mixed_precision}, the mixed precision training brings 27\% memory savings and $1.44\times \sim 2.02\times$ speedup on NVIDIA 2080 Ti and 3090 GPUs with almost no performance drop. 
For activation compression training, 
we conduct experiments of the 2-layer GCN model with 256 hidden units on three datasets for the activation compression training in the CogDL, as shown in Table~\ref{tab:actnn}. 
The performance demonstrates that training GNNs with actnn brings $6.4\times \sim 16\times$ memory savings compared with 32-bit training. The accuracy with actnn (2,3,4-bit) is almost the same with 32-bit training. 

\begin{figure*}[t]
    \centering
    \subfloat[GSpMM on Reddit]{\includegraphics[width=0.24\textwidth]{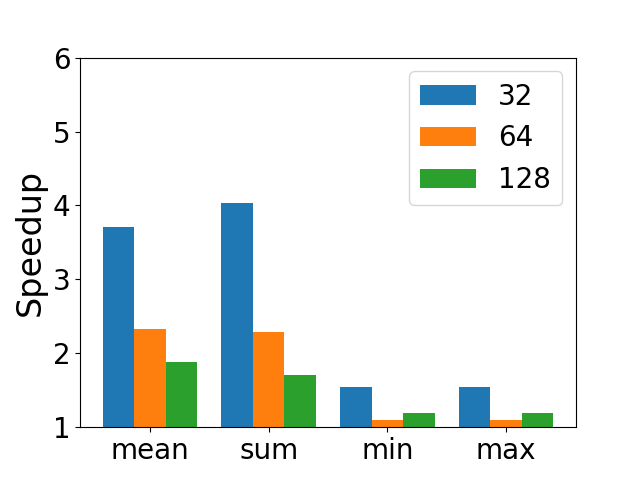}}
	\subfloat[GSpMM on Yelp]{\includegraphics[width=0.24\textwidth]{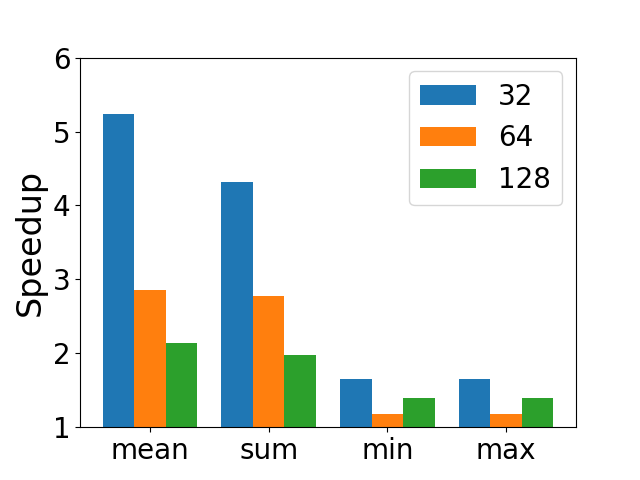}}
    \subfloat[multi-head SpMM on Reddit]{\includegraphics[width=0.24\textwidth]{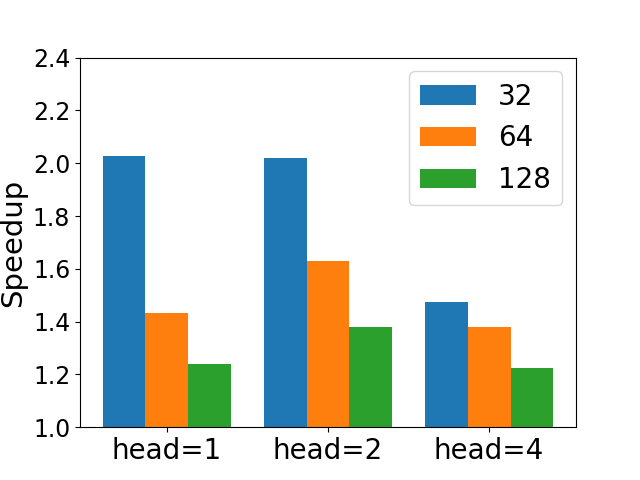}}
	\subfloat[multi-head SpMM on Yelp]{\includegraphics[width=0.24\textwidth]{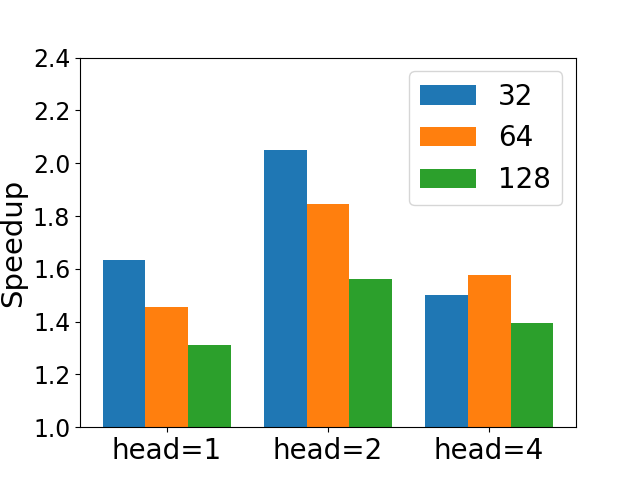}}
    \caption{Speedup of GSpMM and multi-head SpMM with 32, 64, 128 hidden units compared with DGL. (a) $1.70\times \sim 4.04\times$ speedup with mean and sum as reduce functions on Reddit. (b) $1.98\times \sim 5.24\times$ speedup with mean and sum as reduce functions on Yelp. (c) $1.22\times \sim 2.03\times$ speedup with 1, 2, 4 attention heads on Reddit. (d) $1.31\times \sim 2.05\times$ speedup with 1, 2, 4 attention heads on Yelp. }
    \label{fig:gspmm}
\end{figure*}

\subsection{Sparse Operators}
We conduct experiments of GSpMM and multi-head SpMM on Reddit and Yelp datasets compared to DGL, as shown in Figure~\ref{fig:gspmm}. For GSpMM, CogDL achieves $1.17\times \sim 3.47\times$ speedup with different reduce functions on Reddit and $1.27\times \sim 4.01\times$ speedup with different reduce functions on Yelp. 
For multi-head SpMM, CogDL achieves $1.22\times \sim 2.03\times$ and $1.31\times \sim 2.05\times$ speedup on Reddit and Yelp datasets, respectively.

\begin{table}[t!]
    \centering
    \caption{End-to-end inference time in seconds of 2-layer GCN and GAT models with hidden size 128. The GAT model uses 4 attention heads. OOM means out of memory. PyG will be more likely to occur out of memory because it is using the torch-scatter method that will store the temporary edge embedding in the GPU memory.}
    \begin{tabular}{lllccc}
    \toprule
        Model
        & GPU
        & Dataset
        & PyG & DGL & CogDL\\
    \midrule
        \multirow{6}{*}{GCN} 
        & \multirow{3}{*}{2080Ti (11G)} 
        & Flickr & 0.004 & 0.007 & 0.004 \\
        & & Reddit & 0.045 & 0.049 & 0.039 \\
        & & Yelp & 0.053 & 0.063 & 0.042 \\
        \cmidrule{2-6}
        & \multirow{3}{*}{3090 (24G)} 
        & Flickr  & 0.002 & 0.004 & 0.002 \\
        & & Reddit & 0.023 & 0.031 & 0.022 \\
        & & Yelp & 0.029 & 0.036 & 0.023 \\  
    \midrule
        \multirow{6}{*}{GAT} 
        & \multirow{3}{*}{2080Ti (11G)}
        & Flickr & 0.033 & 0.019 & 0.014 \\
        & & Reddit  & OOM   & 0.214 & 0.131 \\
        & & Yelp  & OOM & OOM & 0.138 \\
        \cmidrule{2-6}
        & \multirow{3}{*}{3090 (24G)}
        & Flickr & 0.021 & 0.013 & 0.009 \\
        & & Reddit & OOM   & 0.112 & 0.080  \\
        & & Yelp & OOM   & 0.107 & 0.081 \\
    \bottomrule
    \end{tabular}
    \label{tab:spmm_time}
\end{table}

\subsection{End-to-End Performance}
We compare the end-to-end inference time of GCN and GAT models on several datasets with other popular GNN frameworks: CogDL v0.5.2, PyTorch-Geometric (PyG) v2.0.2, and Deep Graph Library (DGL) v0.7.2 with PyTorch backend. 
The statistics of datasets could be found in Table~\ref{node_dataset}.
We conduct experiments using Python 3.7.10 and PyTorch v1.8.0 on servers with Nvidia GeForce RTX 2080 Ti (11GB GPU Memory) and 3090 (24GB GPU Memory). 
From Table~\ref{tab:spmm_time}, CogDL achieves at most $2\times$ speedup on the 2-layer GCN model compared with PyG and DGL. 
For the 2-layer GAT model, CogDL achieves $1.32\times \sim 2.36\times$ speedup compared with PyG and DGL. Furthermore, OOM occurs when running the PyG's GAT model on Reddit and Yelp datasets, even using 3090 (24G). 
The GAT model implemented by DGL also cannot run the Yelp dataset using 2080Ti (11G). 
The results demonstrate significant advantages of CogDL in inference time and memory savings, compared to state-of-the-art GNN frameworks.

\section{Benchmarking}
\label{sec:exp}

In this section, with CogDL, we introduce several downstream tasks to evaluate implemented methods. 
The statistics of the datasets are shown in Table~\ref{node_dataset} and Table~\ref{graph_dataset} in the Appendix. 
Based on the \texttt{experiment} API, we build a reliable leaderboard for each task, which maintain state-of-the-art results.

\begin{table*}[htbp]
	\caption{Micro-F1 score (\%) of network embedding methods reproduced by CogDL. 50\% of nodes are labeled for training in PPI, Blogcatalog, and Wikipedia, 5\% in DBLP and Flickr. These datasets correspond to different downstream scenarios. }
	\begin{tabular}{llcccccc}
		\toprule
		Rank & Method & PPI (50\%)  & Wikipedia (50\%) & Blogcatalog (50\%) & DBLP (5\%) & Flickr (5\%) & \textit{Reproducible} \\ 
		\midrule
		1 & NetMF~\cite{qiu2018network}    & 23.73 $\pm$ 0.22 & \textbf{57.42 $\pm$ 0.56} & \textbf{42.47 $\pm$ 0.35} & 56.72 $\pm$ 0.14 & 36.27 $\pm$ 0.17 & \textit{Yes} \\
		2 & ProNE~\cite{zhang2019prone}   & \textbf{24.60 $\pm$ 0.39} & 56.06 $\pm$ 0.48 & 41.16 $\pm$ 0.26 & 56.85 $\pm$ 0.28 & \textbf{36.56 $\pm$ 0.11} & \textit{Yes} \\
		3 & NetSMF~\cite{qiu2019netsmf}   & 23.88 $\pm$ 0.35 & 53.81 $\pm$ 0.58 & 40.62 $\pm$ 0.35 & \textbf{59.76 $\pm$ 0.41} & 35.49 $\pm$ 0.07 & \textit{Yes} \\
		4 & Node2vec~\cite{grover2016node2vec} & 20.67 $\pm$ 0.54 & 54.59 $\pm$ 0.51 & 40.16 $\pm$ 0.29 & 57.36 $\pm$ 0.39 & 36.13 $\pm$ 0.13 & \textit{Yes} \\
		5 & LINE~\cite{tang2015line}     & 21.82 $\pm$ 0.56 & 52.46 $\pm$ 0.26 & 38.06 $\pm$ 0.39 & 49.78 $\pm$ 0.37 & 31.61 $\pm$ 0.09 & \textit{Yes} \\
		6 & DeepWalk~\cite{perozzi2014deepwalk} & 20.74 $\pm$ 0.40 & 49.53 $\pm$ 0.54 & 40.48 $\pm$ 0.47 & 57.54 $\pm$ 0.32 & 36.09 $\pm$ 0.10 & \textit{Yes} \\
		7 & SpectralClustering~\cite{tang2011leveraging} & 22.48 $\pm$ 0.30 & 49.35 $\pm$ 0.34 & 41.41 $\pm$ 0.34 & 43.68 $\pm$ 0.58 & 33.09 $\pm$ 0.07 & \textit{Yes} \\
		8 & Hope~\cite{ou2016asymmetric}     & 21.43 $\pm$ 0.32 & 54.04 $\pm$ 0.47 & 33.99 $\pm$ 0.35 & 56.15 $\pm$ 0.22 & 28.97 $\pm$ 0.19 & \textit{Yes} \\
		9 & GraRep~\cite{cao2015grarep}   & 20.60 $\pm$ 0.34 & 54.37 $\pm$ 0.40 & 33.48 $\pm$ 0.30 & 52.76 $\pm$ 0.42 & 31.83 $\pm$ 0.12 & \textit{Yes} \\ 
		\bottomrule
	\end{tabular}
	\label{leader_unc}
\end{table*}

\subsection{Network Embedding}

\vpara{Setup.}
Unsupervised node classification task aims to learn a mapping function $f: V \mapsto \mathbb{R}^d $ that projects each node to a $d$-dimensional space $(d \ll |V|)$ in an unsupervised manner.  Structural properties of the network should be captured by the mapping function. 
We collect the most popular datasets used in the unsupervised node classification task, including 
BlogCatalog~\cite{zafarani2009social},
Wikipedia~\cite{qiu2018network}, 
PPI~\cite{breitkreutz2008biogrid},
DBLP~\cite{tang2008arnetminer},
Flickr~\cite{tang2009relational}. 
We implement and compare the following methods for the unsupervised node classification task, including 
SpectralClustering~\cite{tang2011leveraging},
DeepWalk~\cite{perozzi2014deepwalk},
LINE~\cite{tang2015line},
node2vec~\cite{grover2016node2vec},
GraRep~\cite{cao2015grarep},
HOPE~\cite{ou2016asymmetric},
NetMF~\cite{qiu2018network},
ProNE~\cite{zhang2019prone},
NetSMF~\cite{qiu2019netsmf}. 
Skip-gram network embedding considers the vertex paths traversed by random walks over the network as the sentences and leverages Skip-gram for learning latent vertex representation. For matrix factorization based methods, they first compute a proximity matrix and perform matrix factorization to obtain the embedding. 
Actually, NetMF ~\cite{qiu2018network} has shown the aforementioned Skip-gram models with negative sampling can be unified into the matrix factorization framework with closed forms. 
We run all network embedding methods on five real-world datasets and report the Micro-F1 results (\%) using logistic regression with L2 normalization.

\vpara{Results and Analysis.}
Table~\ref{leader_unc} shows the results and we find some interesting observations. 
DeepWalk and node2vec consider vertex paths traversed by random walk to reach high-order neighbors. NetMF and NetSMF factorize diffusion matrix $\sum_{i=0}^{K} \alpha_i \mA^i$ rather than adjacency matrix $\mA$. ProNE and LINE are essentially 1-order methods, but ProNE further propagates the embeddings to enlarge the receptive field. Incorporating global information can improve performance but may hurt efficiency. The propagation in ProNE, which is similar to graph convolution, shows that incorporating global information as a post-operation is effective. 
The results demonstrate that matrix factorization (MF) methods like NetMF and ProNE are very powerful and full of vitality as they outperform Skip-gram based methods in almost all datasets. ProNE and NetSMF are also of high efficiency and scalability and able to embed super-large graphs in feasible time in one single machine.

\begin{table}[htbp]
	\caption{Accuracy (\%) reproduced by CogDL for semi-supervised and self-supervised node classification on Citation datasets. $\downarrow$ and $\uparrow$ mean our results are lower or higher than the result in original papers. }
	\begin{tabular}{lcccc}
		\toprule
		Method	    &	Cora	    &	Citeseer	&	Pubmed        & \textit{Reproducible} \\ 
		\midrule
		GRAND~\cite{feng2020grand}	            &	84.8	&	\textbf{75.1}	&	\textbf{82.4}    & \textit{Yes}\\
		GCNII~\cite{chen2020gcnii}	            &	\textbf{85.1}	&	71.3	&	80.2    & \textit{Yes} \\
		MVGRL~\cite{hassani2020mvgrl}	        &	83.6 $\downarrow$	&	73.0	&	80.1    & \textit{Partial}\\
		APPNP~\cite{klicpera2019appnp}	        &	84.3 $\uparrow$	&	72.0	&	80.0    & \textit{Yes}\\
		Graph-Unet~\cite{gao2019graph}          &   83.3 $\downarrow$    &   71.2 $\downarrow$    &   79.0    & \textit{Partial} \\
		GDC~\cite{klicpera2019diffusion}        &   82.5    &   72.1    &   79.8    & \textit{Yes} \\
		GAT~\cite{Velickovic:18GAT}		        &	82.9	&	71.0	&	78.9    & \textit{Yes} \\
		DropEdge~\cite{rong2019dropedge}        &   82.1    &   72.1    &   79.7    & \textit{Yes} \\   
		GCN~\cite{kipf2016semi}	    	        &	81.5	&	71.4 $\uparrow$	&	79.5    & \textit{Yes} \\
		DGI~\cite{velickovic2019dgi}		    &	82.0	&	71.2	&	76.5    & \textit{Yes}\\
		JK-net~\cite{xu2018representation}     &   81.8    &   69.5    &   77.7    & \textit{Yes} \\
		Chebyshev~\cite{defferrard2016chebynet}	&	79.0	&	69.8	&	68.6    & \textit{Yes} \\
		\bottomrule
	\end{tabular}
	\normalsize
	\label{leader_snc}
\end{table}

\begin{table}[htbp]
    \centering
    \caption{Results of node classification on fully-supervised datasets, including full-batch and sampling-based methods. Flickr, Reddit, and ogbn-arxiv use accuracy metric, whereas PPI uses Micro-F1 metric.}
    \begin{tabular}{lcccc}
        \toprule
                    & Flickr             & PPI                 & Reddit              & arxiv\\
        \midrule
        GCNII       &  \textbf{52.6$\pm$0.1}    & 96.5$\pm$0.2     &  \textbf{96.4$\pm$0.0}     & \textbf{72.5$\pm$0.1} \\
        GraphSAINT  &  52.0$\pm$0.1    & \textbf{99.3$\pm$0.1}        &  96.1$\pm$0.0     & 71.5$\pm$0.2 \\
        GAT &  51.9$\pm$0.3 & 97.3$\pm$0.2  & 95.9$\pm$0.1 & 72.3$\pm$0.1	\\
        Cluster-SAGE & 50.9$\pm$0.1 & 97.9$\pm$0.1  & 95.8$\pm$0.1 & 69.8$\pm$0.1 \\ 
        GCN         &  52.4$\pm$0.1    & 75.7$\pm$0.1    &  95.1$\pm$0.0     & 	71.7$\pm$0.3        \\
        APPNP       &  52.3$\pm$0.1   & 62.3$\pm$0.1     &  96.2$\pm$0.1     & 70.7$\pm$0.1        \\
        PPRGo       &  51.1$\pm$0.2   & 48.6$\pm$0.1     &  94.5$\pm$0.1     & 69.2$\pm$0.1 \\
        SGC  & 50.2$\pm$0.1 & 47.2$\pm$0.1 & 94.5$\pm$0.1 & 67.1$\pm$0.0 \\
        \bottomrule
    \end{tabular}
    \normalsize
    \label{tab:lead_nc_ind}
\end{table}

\subsection{Node Classification}

\vpara{Setup.} 
This task is for node classification with GNNs in semi-supervised and self-supervised settings. Different from the previous part, nodes in these graphs, like Cora and Reddit, have node features and are fed into GNNs with prediction or representation as output. Cross-entropy loss and contrastive loss are set for semi-supervised and self-supervised settings, respectively. 
The datasets consist of two parts, including both semi-supervised and fully-supervised settings. 
Semi-supervised datasets include three citation networks, Citeseer, Cora, and Pubmed~\cite{sen2008collective}. 
Fully-supervised datasets include social networks (Reddit, Yelp, and Flickr), bioinformatics (PPI) from GraphSAINT~\cite{graphsaint-iclr20}, and ogbn-arxiv~\cite{hu2020open}. 
We implement all the following models, including 
ChebyNet~\cite{defferrard2016chebynet},
GCN~\cite{kipf2016semi},
GAT~\cite{Velickovic:18GAT},
GraphSAGE~\cite{hamilton2017inductive},
APPNP~\cite{klicpera2019appnp},
DGI~\cite{velickovic2019dgi},
GCNII~\cite{chen2020gcnii},
MVGRL~\cite{hassani2020mvgrl},
GRAND~\cite{feng2020grand},
DropEdge~\cite{rong2019dropedge},
Graph-Unet~\cite{gao2019graph},
GDC~\cite{klicpera2019diffusion},
PPRGo~\cite{bojchevski2020scaling},
GraphSAINT~\cite{graphsaint-iclr20}. 
For evaluation, we use prediction accuracy for multi-class and micro-F1 for multi-label datasets.

\begin{table*}[htbp]
    \centering
    \caption{Accuracy Results (\%) of both unsupervised and supervised graph classification. $\downarrow$ and $\uparrow$ mean our results are lower or higher than the results in original papers. $\ddagger$ means the experiment is not finished in 24 hours for one seed. }
	\begin{tabular}{lccccccccc} 
		\toprule
		Algorithm   & MUTAG & PTC   & NCI1  &PROTEINS & IMDB-B & IMDB-M & COLLAB & REDDIT-B & \textit{Reproducible} \\ 
		\midrule
		GIN~\cite{xu2019gin}         & \textbf{92.06} & \textbf{67.82} & \textbf{81.66} & 75.19   & \textbf{76.10}  & 51.80  & 79.52  & 83.10 $\downarrow$ & \textit{Yes}\\
		InfoGraph~\cite{sun2019infograph}   & 88.95 & 60.74 & 76.64 & 73.93   & 74.50  & 51.33  & 79.40  & 76.55 & \textit{Yes} \\
		graph2vec~\cite{narayanan2017graph2vec}   & 83.68 & 54.76 $\downarrow$ & 71.85 & 73.30   & 73.90  & \textbf{52.27}  & \textbf{85.58} $\uparrow$  & \textbf{91.77} & \textit{Yes}\\
		SortPool~\cite{zhang2018sortpool}    & 87.25 & 62.04 & 73.99 $\uparrow$ & 74.48   & 75.40  & 50.47  & 80.07 $\uparrow$  & 78.15 & \textit{Yes} \\
		DiffPool~\cite{ying2018diffpool}    & 85.18 & 58.00 & 69.09 & 75.30   & 72.50  & 50.50  & 79.27  & 81.20 & \textit{Yes} \\
		PATCHY\_SAN~\cite{niepert2016learning} & 86.12 & 61.60 & 69.82 & \textbf{75.38}   & 76.00 $\uparrow$  & 46.40  & 74.34  & 60.61 & \textit{Yes}\\
		DGCNN~\cite{wang2019dgcnn}       & 83.33 & 56.72 & 65.96 & 66.75   & 71.60  & 49.20  & 77.45  & 86.20 & \textit{Yes} \\
		SAGPool~\cite{lee2019sagpool}     & 71.73 $\downarrow$ & 59.92 & 72.87 & 74.03   & 74.80  & 51.33  &  \_\_$\ddagger$     & 89.21 & \textit{Yes} \\
		DGK~\cite{yanardag2015dgk}         & 85.58 & 57.28 & \_\_$\ddagger$     & 72.59   & 55.00 $\downarrow$  & 40.40 $\downarrow$  & \_\_$\ddagger$      & \_\_$\ddagger$     & \textit{Partial}\\
		\bottomrule
	\end{tabular}
	\label{leader_gc}
\end{table*}

\vpara{Results and Analysis.}
Table~\ref{leader_snc} and Table~\ref{tab:lead_nc_ind} summarize the evaluation results of all compared models in semi-supervised and fully-supervised datasets, respectively, under the setting of node classification. 
We observe that incorporating high-order information plays an important role in improving the performance of models,
especially in citation datasets (Cora, CiteSeer, and PubMed) that are of relatively small scale. 
Most high-order models, such as GRAND, APPNP, and GDC, aim to use graph diffusion matrix $\bar{\mA}=\sum_{i=0}^{K}\alpha_i \hat{\mA}^i$ to collect information of distant neighbors. 
GCNII uses the residual connection with identity mapping to resolve over-smoothing in GNNs and achieves remarkable results.  
GRAND~\cite{feng2020grand} leverages unlabelled data in semi-supervised settings and obtains state-of-the-art results. 
As for self-supervised methods, DGI and MVGRL maximize local and global mutual information. MVGRL performs better by replacing the adjacency matrix with graph diffusion matrix but is less scalable. On the contrary, GRACE, which optimizes InfoNCE, does not perform well on citation datasets with the public split.

\subsection{Graph Classification}
\vpara{Setup.}
Graph classification assigns a label to each graph and aims to map graphs into vector spaces. 
Graph kernels are historically dominant and employ a kernel function to measure the similarity between pairs of graphs to map graphs into vector spaces with deterministic mapping functions. But they suffer from computational bottlenecks. 
Recently, graph neural networks attract much attention and indeed show promising results in this task. In the context of graph classification, GNNs often employ the readout operation to obtain a compact representation of the graph and apply classification based on the readout representation. 
We collect eight popular benchmarks used in graph classification tasks, including Bioinformatics datasets (PROTEINS, MUTAG, PTC, and NCI1), and Social networks (IMDB-B, IMDB-M, COLLAB, REDDIT-B). 
We implement the following graph classification models and compare their results: 
GIN~\cite{xu2019gin}, DiffPool~\cite{ying2018diffpool}, SAGPool~\cite{lee2019sagpool}, SortPool~\cite{zhang2018sortpool}, DGCNN~\cite{wang2019dgcnn} and Infograph~\cite{sun2019infograph} are based on GNNs. PATCHY\_SAN~\cite{niepert2016learning} is inspired by convolutional neural networks. Deep graph kernels (DGK)~\cite{yanardag2015dgk} and graph2vec~\cite{narayanan2017graph2vec} use graph kernels. 
As for evaluation, for supervised methods we adopt 10-fold cross-validation with 90\%/10\% split and repeat 10 times; for unsupervised methods, we perform the 10-fold cross-validation with LIB-SVM~\cite{CC01a}. Then we report the accuracy for classification performance of all methods on these datasets.

\vpara{Results and Analysis.} 
Table~\ref{leader_gc} reports the results of the aforementioned models on the task, including both unsupervised and supervised graph classification. 
The development of GNNs for graph classification is mainly in two aspects. One line (like GIN) aims to design more powerful convolution operations to improve the expressiveness. Another line is to develop effective pooling methods to generate the graph representation. 
Neural network based methods show promising results in bioinformatics datasets (MUTAG, PTC, PROTEINS, and NCI1), where nodes are with given features. But in social networks (IMDB-B, IMDB-M, COLLAB, REDDIT-B) lacking node features, methods based on graph kernels achieve good performance and even surpass neural networks. Graph kernels are more capable of capturing structural information to discriminate non-isomorphic graphs, while GNNs are better encoders with features. 
Global pooling, which is used in GIN and SortPool, collects node features and applies a readout function. Hierarchical pooling, such as DiffPool and SAGPool, is proposed to capture structural information in different graph levels, including nodes and subgraphs. The experimental results indicate that, though hierarchical pooling seems more complex and intuitively would capture more information, it does not show significant advantages over global pooling.

\subsection{Other Graph Tasks}

As shown in Table~\ref{tab:all_tasks}, CogDL also provides other important tasks in graph learning. In this section, we briefly introduce the setting and evaluation of some tasks, including heterogeneous node classification, link prediction, multiplex link prediction, etc. 

\vpara{Heterogeneous Node Classification.}
This task is built for heterogeneous GNNs conducted on heterogeneous graphs such as academic graphs. 
For heterogeneous node classification, we use macro-F1 to evaluate the performance of all heterogeneous models under the setting of Graph Transformer Networks (GTN)~\cite{yun2019gtn}. The heterogeneous datasets include DBLP, ACM, and IMDB. 
We implement these models, including GTN~\cite{yun2019gtn}, HAN~\cite{wang2019han}, PTE~\cite{tang2015pte}, metapath2vec~\cite{dong2017metapath2vec}, hin2vec~\cite{fu2017hin2vec}.

\vpara{Link Prediction.}
Many real applications, such as citation link prediction, can be viewed as a link prediction task~\cite{kipf2016variational}. 
We formulate the link prediction task as: given a part of edges in the graph, after learning representations for each vertex $v\in V$, the method needs to accurately predict the edges that belong to the rest of the edges rather than random edges. 
In practice, we remove 15 percent of original edges for each dataset and adopt ROC AUC~\cite{hanley1982meaning} as the evaluation metric. 

\vpara{Multiplex Link Prediction.}
GATNE~\cite{cen2019representation} formalizes the representation learning problem of multiplex networks and proposes a unified framework to solve the problem in transductive and inductive settings. We follow the setting of GATNE~\cite{cen2019representation} to build the multiplex heterogeneous link prediction task. 
In CogDL, for those methods that only deal with homogeneous networks, we separate the graphs based on the edge types and train the method on each separate graph. 
We also adopt ROC AUC as the evaluation metric.

\vpara{Knowledge Graph Completion.}
This task aims to predict missing links in a knowledge graph through knowledge graph embedding. In CogDL, we implement two families of knowledge graph embedding algorithms: triplet-based knowledge graph embedding and knowledge-based GNNs. The former includes TransE \cite{bordes2013translating},  DistMult \cite{Yang2015EmbeddingEA}, ComplEx \cite{trouillon2016complex}, and RotatE \cite{sun2019rotate}; the latter includes RGCN \cite{schlichtkrull2018modeling} and CompGCN \cite{vashishth2019compositionbased}. We evaluate all implemented algorithms with standard benchmarks including FB15k-237, WN18 and WN18RR with Mean Reciprocal Rank (MRR) metric.

\section{Discussions}
\label{sec:discuss}

With the success of graph neural networks, there are dozens of graph libraries developed in the past few years, including 
PGL~\cite{pgl}, Graph Nets~\cite{battaglia2018relational}, Jraph~\cite{jraph2020github}, Graph-Learn~\cite{zhu2019aligraph}, Spektral~\cite{grattarola2020graph}, StellarGraph~\cite{StellarGraph}, Deep Graph Library (DGL)~\cite{wang2019deep}, and PyTorch Geometric (PyG)~\cite{fey2019fast}. 
Among them, PyG~\cite{fey2019fast} and DGL~\cite{wang2019deep} are two of the most well-known libraries with efficient operators and easy-to-use APIs. 
PyG provides CUDA kernels for sparse operators with high data throughput. 
PyG 2.0 further utilizes GraphGym~\cite{you2020design} to achieve new features like reproducible configuration and hyperparameter optimization. 
DGL not only provides flexible and efficient message-passing APIs but also allows users to easily port and leverage the existing components across multiple deep learning frameworks (e.g., PyTorch~\cite{paszke2019pytorch}, TensorFlow~\cite{tensorflow2015-whitepaper}). 
However, CogDL not only supports flexible implementations of GNN models but also provides a unified trainer for the end-to-end training of GNN models with efficient techniques.

\section{Conclusions}
\label{sec:conclusion}

This paper introduces CogDL, a comprehensive library for graph deep learning that focuses on the research and applications of graph representation learning.   
CogDL provides simple APIs such that users can train and evaluate graph learning methods with ease and efficiency.  
Besides, CogDL implements many state-of-the-art models with wrappers for various graph tasks. 
Finally, CogDL collects and maintains leaderboards with reproducible configurations on widely-used public datasets. 
In the future, we will consistently improve the efficiency of CogDL and support more features to satisfy the latest need of both scientific and industrial applications. 
For example, the dynamic scenario of graph learning, where nodes and edges will continuously be added and removed, needs to be considered. 
Besides, graph self-supervised learning and pre-training have become one of the most exciting research directions in recent years, which presents a new challenge for graph libraries.

\begin{acks}
This research was supported by the Natural Science Foundation of China (NSFC) 61825602 and 62276148 and the Tsinghua-Siemens Joint Research Center for Industrial Intelligence and Internet of Things (JCIIOT).
\end{acks}


\bibliographystyle{ACM-Reference-Format}
\bibliography{reference}

\clearpage

\appendix
\begin{figure*}
    \centering
    \includegraphics[width=0.99\textwidth]{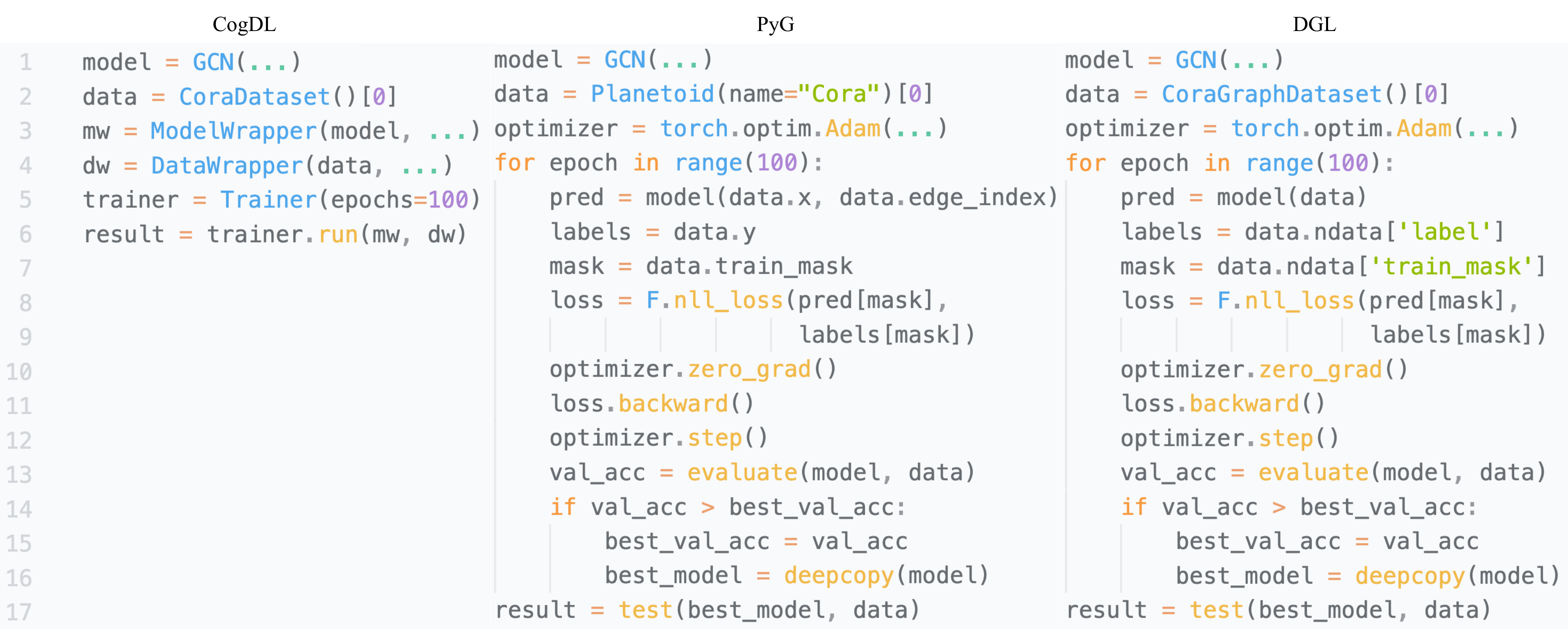}
    \caption{Illustration of training the GCN model on the Cora dataset using CogDL, PyG, and DGL. The trainer of CogDL takes over the entire loop of the training and evaluation process. The ModelWrapper consists of training and testing steps, while the DataWrapper handles the data preparation and transformation. }
    \label{fig:demo}
\end{figure*}

\begin{table*}[htbp]
    \centering
    \caption{Dataset statistics for node classification (``m'' stands for multi-label classification)}
	\begin{tabular}{llrrrcr}
		\toprule
		Setting & Dataset & \#Nodes & \#Edges & \#Features & \#Classes & \# Train / Val / Test \\
		\midrule
		\multirow{3}{*}{Semi-supervised Setting} & Cora & 2,708 & 5,429 & 1,433 & 7 & 140 / 500 / 1,000 \\
		 & Citeseer & 3,327 & 4,732 & 3,703 & 6 & 120 / 500 / 1,000 \\
		 & Pubmed & 19,717 & 44,338 & 500 & 3 & 60 / 500 / 1,000 \\
		\midrule
		\multirow{5}{*}{Fully-Supervised Setting} 
		 & PPI & 56,944 & 818,736 & 50 & 121 (m) & 0.79 / 0.11 / 0.10 \\ 
		 & Flickr & 89,350 & 899,756 & 500 & 7 & 0.50 / 0.25 / 0.25 \\
		 & Reddit & 232,965 & 11,606,919 & 602 & 41 & 0.66 / 0.10 / 0.24 \\
		 & Yelp   & 716,847 & 6,977,410 & 300 & 100 (m) & 0.75 / 0.10 / 0.15 \\
		 & ogbn-arxiv & 169,343 & 1,166,243 & 128 & 40 & 0.54 / 0.18 / 0.28 \\
		\midrule
		\multirow{5}{*}{Network Embedding Setting} 
		 & PPI & 3,890 & 76,584 & - & 50 (m) & 0.50 / - / 0.50 \\
		 & Wikipedia & 4,777 & 184,812 & - & 40 (m) & 0.50 / - / 0.50 \\
		 & Blogcatalog & 10,312 & 333,983 & - & 39 (m) & 0.50 / - / 0.50 \\
		 & DBLP & 51,264 & 127,968 & - & 60 (m) & 0.05 / - / 0.95 \\
		 & Flickr & 80,513 & 5,899,882 & - & 195 (m) & 0.05 / - / 0.95 \\
		\bottomrule
	\end{tabular}
	\label{node_dataset}
\end{table*}

\begin{table}[htbp]
    \centering
    \caption{Dataset statistics for graph classification}
	\begin{tabular}{lrrrrr}
		\toprule
		Dataset & \#Graphs & \#Classes & \#Feats & \#Nodes & \#Edges \\
		\midrule
		MUTAG & 188 & 2 & 7 & 17.9 & 19.8 \\
		PTC & 344 & 2 & 18 & 14.3 & 14.7 \\
		PROTEINS & 1,113 & 2 & 3 & 39.1 & 72.8 \\
		NCI1 & 4,110 & 2 & 37 & 29.8 & 32.3 \\
		\midrule
		IMDB-B & 1,000 & 2 & - & 19.8 & 96.5 \\
		IMDB-M & 1,500 & 3 & - & 13.0 & 65.9 \\
		REDDIT-B & 2,000 & 2 & - & 429.6 & 497.8 \\
		COLLAB & 5,000 & 3 & - & 74.5 & 2457.8 \\
		\bottomrule
	\end{tabular}
	\label{graph_dataset}
\end{table}

\section{CogDL Package}
In this section, we introduce the details of the CogDL library. 

\subsection{Basic Components}
The basic elements in CogDL include \texttt{Graph}, \texttt{Dataset}, and \texttt{Model}. 

\vpara{Graph.}
The \texttt{Graph} is the basic data structure of CogDL to store graph data with abundant graph operations. \texttt{Graph} supports both convenient graph modification, such as removing/adding edges, and efficient computing, such as SpMM. We also provide common graph manipulations, including 
adding self-loops, graph normalization, sampling neighbors, obtaining induced subgraphs, etc. 

\vpara{Dataset.}
The \texttt{dataset} component reads in data and processes it to produce tensors of appropriate types. Each dataset specifies its loss function and evaluator through function \texttt{get\_loss\_fn} and \texttt{get\_evaluator} for training and evaluation. 
Our package provides many real-world datasets and easy-to-use interfaces for users to define customized datasets. 
CogDL provides two ways to construct a customized dataset. One way is to convert raw data files into the required format of CogDL and specify the argument $data\_path$. The other way is to define a customized dataset class with the necessary functions to load and process the data.

\begin{lstlisting}[language=Python,breaklines=True,frame=single,numbers=none,caption=Definition of datasets,label={lst:dataset}]
class CoraDataset(Dataset):
  def process(self):
    """preprocess data to cogdl.Graph"""
  def get_evaluator(self):
    """Return evaluator defined in CogDL"""
    return accuracy
  def get_loss_fn(self):
    """Return loss function defined in CogDL"""
    return cross_entropy_loss
\end{lstlisting}

\vpara{Model.}
CogDL implements various graph neural networks as research baselines and for applications. A GNN layer such as \texttt{GCNLayer} has already been implemented based on \texttt{Graph} and efficient sparse operators. A \texttt{model} in the package consists of the model builder (i.e., \texttt{\_\_init\_\_}) and forward propagation (i.e., \texttt{forward}) functions in the PyTorch style. 
Listing~\ref{lst:model} shows the APIs implemented in each model to provide a unified paradigm for usage, where \texttt{add\_args} is used to define model-specific hyperparameters for experiments.

\noindent
\begin{lstlisting}[language=Python,breaklines=True,frame=single,numbers=none,caption=Definition of GNN models,label={lst:model}]
from cogdl.layers imoprt GCNLayer

class GCN(BaseModel):
  def add_args(parser):
    parser.add_argument("--hidden", type=int)
  
  def __init__(self, nfeats, nclasses, hidden):
    self.layer1 = GCNLayer(nfeats, hidden)
    self.layer2 = GCNLayer(hidden, nclasses)
  
  def forward(self, graph):
    h = graph.x
    h = self.layer1(graph, h)
    h = self.layer2(graph, h)
    return h
\end{lstlisting}

\subsection{Basic Usage}
\label{subsec:basic_usage}

Figure~\ref{fig:demo} illustrates the benefits of CogDL's unified trainer and modular wrappers over PyG and DGL. 
Users can directly use the \texttt{Trainer} for running the experiments and do not need to write tedious code to build the training loop for GNN training and evaluation. 
Besides, we introduce another easy-to-use usage for conducting experiments through \texttt{experiment} API. 
We can feed a dataset, a model, and hyper-parameters to the \texttt{experiment} API, which calls the \texttt{Trainer} API. Producing the results with one-line code provides a convenient way for experiments. Furthermore, we integrate a popular library, Optuna~\cite{akiba2019optuna}, into CogDL to enable hyper-parameter search. 
By feeding the \texttt{search\_space} function that defines the search space of hyper-parameters, \revise{CogDL will start the searching process for the validation metric and output the best results.}

\noindent
\begin{lstlisting}[language=Python,breaklines=true,frame=single,numbers=none,caption=Basic usages of CogDL]
from cogdl import experiment
experiment(dataset="cora", model="gcn", hidden_size=32, max_epoch=200) # basic usage

def func(trial):
  return {
    "hidden_size": trial.suggest_categorical("hidden_size", [32, 64, 128]),
    "dropout": trial.suggest_uniform("dropout", 0.1, 0.8),
  }
experiment(dataset="cora", model="gcn", search_space=func) # hyper-parameter search
\end{lstlisting}

\section{Applications}
\label{sec:app}

\begin{figure}[thbp]
    \centering
    \includegraphics[width=0.47\textwidth]{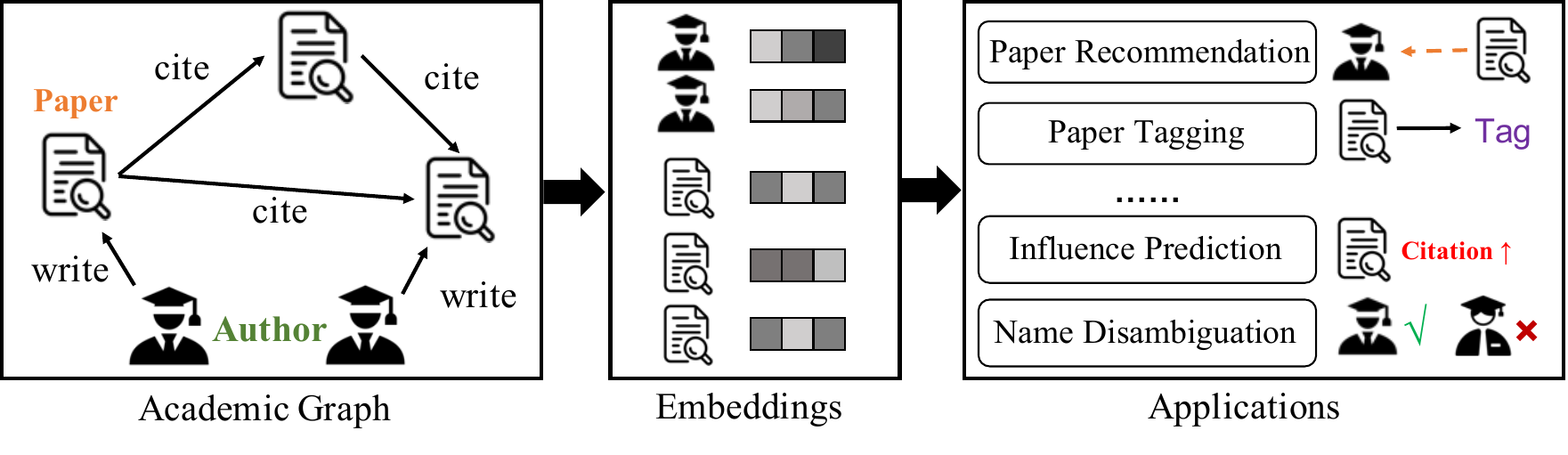}
    \caption{Illustration of how CogDL helps real-world applications in AMiner. Most applications of AMiner rely on a large-scale academic graph. CogDL can help AMiner to generate accurate embeddings for papers and scholars from the billion-scale academic graph for downstream tasks. }
    \label{fig:application}
\end{figure}

Our library has been successfully applied to a large-scale academic mining and search system---AMiner (\url{https://www.aminer.cn/} )---for the analysis of papers and scholars, as illustrated in Figure~\ref{fig:application}. 
There are more than 320 million papers with more than 1 billion citation links in the AMiner database. 
Most applications (e.g., paper tagging) in AMiner are related to academic papers and scholars. 
Therefore, it is important to obtain accurate embeddings for papers and scholars for downstream tasks. 

Take paper tagging as an example. 
Each publication in AMiner has several tags extracted from the raw text (e.g., title and abstract) of each publication. 
However, publications with citation links may have similar tags, and we can utilize the citation network to improve the quality of tags.
The paper tagging problem can be considered as a multi-label node classification task, where each label represents a tag. 
We can utilize paper embeddings to improve the quality of paper tags since we find many papers missing tags in the AMiner database. With the help of the AMiner team, we train a downstream classifier for the paper tagging task on 4.8 million papers in computer science. The experimental results demonstrate that those embeddings could help improve the recall by 12.8\% compared with the previous method. 
In addition to the paper tagging task, the embeddings generated by CogDL can also help the paper recommendation of AMiner. 
Specifically, the AMiner homepage gives paper recommendations for users based on users' historical behaviors. 
The quality of the paper embeddings plays an important role in the recommendation. 
We hope that our library can help more users to build better services, especially those web-related applications, benefiting the community of graph learning.

\end{document}